# WORK PROBABILITY DISTRIBUTION FOR A FERROMAGNET WITH LONG-RANGED AND SHORT- RANGED CORRELATIONS

J. K. Bhattacharjee[1,2], T. R. Kirkpatrick[1] and J. V. Sengers[1]

[1] Institute for Physical Science and Technology, University of Maryland, College Park, Maryland 20742, USA

[2] Department of Theoretical Physics, Indian Association For the Cultivation of Science, Jadavpur, Kolkata 700032, India

ABSTRACT

Work fluctuations and work probability distributions are fundamentally different in systems with short- ranged versus long-ranged correlations. Specifically, in systems with long-ranged correlations the work distribution is extraordinarily broad compared to systems with short-ranged correlations. This difference profoundly affects the possible applicability of fluctuation theorems like the Jarzynski fluctuation theorem. The Heisenberg ferromagnet , well below its Curie temperature, is a system with long-ranged correlations in very low magnetic fields due to the presence of Goldstone modes. As the magnetic field is increased the correlations gradually become short-ranged. Hence, such a ferromagnet is an ideal system for elucidating the changes of the work probability distribution as one goes from a domain with long-ranged correlations to a domain with short-ranged correlations by tuning the magnetic field. A quantitative analysis of this crossover behaviour of the work probability distribution and the associated fluctuations is presented.

I  INTRODUCTION

One of the most significant developments in non-equilibrium statistical mechanics in the last few decades has been the emergence of a set of fluctuation- dissipation theorems [1-6], some of which [4-6] concern the fluctuations in the work done as a system moves from an initial thermodynamic state at temperature $T$ to a final thermodynamic state at the same temperature. The passage from the initial to the final state can occur along any one of the infinitely available paths. Hence one needs to average over an ensemble of paths, when one considers any function of the work ($W$) done in the transition from the initial to the final state. A primary example of such a fluctuation- dissipation theorem is Jarzynski's equality [4] which yields

$$<e^{-\frac{W}{k_BT}}>=e^{-\frac{\Delta F}{k_BT}} \qquad (1.1)$$

where $W$ is the work done in going from state 1 ( initial ) to state 2 ( final), $k_B$ is Boltzmann's constant and where $\Delta F = F_2 - F_1$, with $F_1$ and $F_2$ being the thermodynamic free energies ( Helmholtz energies) associated with states 1 and 2, respectively.

The appearance of an expectation value on the left- hand side of Eq. (1.1) raises questions about the probability distribution of the work fluctuations. The fact that rare events can play a role in determining the tail of the distribution has motivated researchers to find the distribution both theoretically and experimentally [7-11]. Crooks and Jarzynski [12] have considered a particularly illuminating example involving the adiabatic compression of an ideal monatomic gas from an initial volume $V_0$ to a final volume $V_1$. In terms of a parameter $\alpha$ defined by the relation $1+\alpha = (V_0/V_1)^{2/3}$; they found for the work probability distribution $\rho(W)$ in this case ( note that $W > 0$ in this case )

$$\rho(W) = \frac{1}{\alpha k_B T \Gamma(3N/2)} \left(\frac{W}{\alpha k_B T}\right)^{\frac{3N}{2}-1} e^{-\frac{W}{\alpha k_B T}} \qquad (1.2)$$

where $N$ is the number of molecules of the gas.

As noted recently by Kirkpatrick *et al.* [13], this distribution is centered at $W_0 = 3N/2\alpha$ and around the centre can be approximated by the Gaussian form

$$\rho_{\text{approx}}(W) = C e^{-(W-W_0)^2/3N\alpha^2} \qquad (1.3)$$

With the distribution peaked at $W_0$ which is $O(N)$ and with a width which is $O(\sqrt{N})$, $\rho(W)$ is sharply peaked at its maximum for $N \gg 1$. Around the maximum of the distribution, the function $\Omega = \exp(-W/k_B T)$ is already very small and hence fluctuations in it can be significant. The root- mean- square fluctuation $\varepsilon_\Omega$ of $\Omega$ is defined as

$$\varepsilon_\Omega = \frac{\langle \Omega^2 \rangle}{\langle \Omega \rangle^2} - 1 = \frac{\langle \exp(-2W/k_B T) \rangle}{\langle \exp-W/k_B T \rangle^2} - 1 \qquad (1.4)$$

Using the distribution, given by Eq. (1.2), we get.

$$\varepsilon_\Omega = \exp[\frac{3N}{2}\ln(1 + \frac{\alpha^2}{1+2\alpha})] \qquad (1.5)$$

What is striking about the above equation is that the fluctuation is exponentially large in the system size $N$ and thus makes the usefulness of the equality shown in Eq. (1.1) questionable. This will become a problem for large systems where $N \gg 1$ and $\varepsilon_\Omega$ is supposed to be vanishingly small. Instead we have in this case exponentially large ( in the system size ) fluctuations around the mean. Even for systems where $N$ is of $O(1)$, the fluctuations are of the same order as the mean, which makes the approach to the mean value difficult. The Gaussian approximation of Eq. (1.4) yields $\varepsilon_\Omega = \exp[3N\alpha^2/2]$, which agrees with the result obtained from the exact distribution for $\alpha \ll 1$ as expected. These very large fluctuations result from the short-range nature of the correlations. This feature can also be seen in a situation with a set of independent oscillators envisaged by Hijar and Ortiz de Zárate[14].

What if the correlations are of a long-ranged nature ? As shown in a series of papers in the case of a fluid under a fixed temperature gradient [15-18], one has a non-equilibrium steady state (NESS) with generic long-range correlations. These long-ranged correlation are capable of producing large Casimir type forces in confined fluid layers [19-22]. Recently it was found that the work probability distribution in such cases can be significantly different [13]. This prompted us to look at systems with long-ranged correlations in both equilibrium and non-equilibrium situations [23]. A natural candidate for an equilibrium system was a Heisenberg ferromagnet, well below the Curie temperature, in an infinitesimal magnetic field, where the transverse magnetization fluctuations are the Goldstone modes of the system. These modes, being massless, lead to the fluctuations having long-ranged correlations. As anticipated, the long-ranged correlation between the fluctuations lead to a work probability distribution similar to the case of NESS with generic long-ranged fluctuation correlations. The important point about the probability distributions in the case of systems with these long-ranged correlations is that the distribution is wider by orders of magnitude and this greatly reduces the fluctuations $\varepsilon_\Omega$ around $\langle \exp-(W/k_BT) \rangle$.

The Heisenberg ferromagnet at temperatures well below the Curie temperature actually provides a single system where both short-ranged and long-ranged correlations can be probed. Specifically, one can use the external magnetic field as a tuning parameter to go from long-ranged to short-ranged fluctuations. Consequently, this is an ideal testing ground for studying the crossover in the probability distribution of the work and in the fluctuation $\varepsilon_\Omega$, as one goes from long-ranged correlations to short-ranged ones.

In three dimensional space the correlation between the local transverse magnetization fluctuations is long ranged in infinitesimal external fields and falls off as $1/r$, where $r$ is the separation between the local fluctuations. As the magnetic field $h$ is increased this long-ranged correlation is shielded by a screening term of the form $\exp(-r/\xi)$, where $\xi$ is a correlation length which is inversely proportional to $\sqrt{h}$. Consequently, by increasing the magnetic field one should be able to see the work probability distribution become sharper and sharper. Thus ferromagnets are ideal candidates for studying various fine points about the fluctuations defined in Eq. (1.4).

Specifically we will consider a slab of ferromagnetic material where the extensions ( linear dimension $L_\perp$ ) in the x-y plane will be taken to be large while the extension in the z - direction ( also the direction of the external magnetic field ) is a smaller length $L$ . We will define the dimensionless area $A$ as $A = L_\perp^2 / L^2$. This dimensionless area $A$ can be large if $L_\perp \gg L$ but can also be of $O(1)$ when the two lengths are comparable. The comparison between $L$ and the correlation length $\xi$ will determine whether the system is long-ranged ( long- ranged correlation implies $\xi \gg L$ ) or not. We will see that the transition from long-ranged to short-ranged correlations will be governed by a dimensionless parameter $L_h^2 = hL^2 / MJ$ where, $M$ is the magnetization and $J$ is the exchange coupling between the local magnetic moments . For large values of $L_h^2$ ( short- ranged correlations ), it will be seen that the probability distribution is very sharp and centered around $h^{3/2}$. This will be analogous to the short-ranged cases discussed above and one will find large fluctuations around the quantity $\exp(-W / k_B T)$, similar to the result shown in Eq. (1.5). For small values of $L_h^2$ , the distribution will become broad reducing considerably the fluctuations $\varepsilon_\Omega$ as defined in Eq. (1.4).

The primary result of this work is that

$$\varepsilon_\Omega \propto \exp(Af(L_h)) \qquad (1.6)$$

where $f(x)$ is a crossover function . For infinitesimal magnetic fields when correlations are long- ranged and $L_h \ll 1$, $f(L_h) \propto L_h^4$, while for larger magnetic fields , when correlations are short- ranged and $L_h \gg 1$, we find $f(L_h) \propto L_h^3$. In the latter case we are back to Eq. (1.5) and the usefulness of the equality in Eq. (1.1) becomes restricted to single molecules and nano scales , while in the former case we have a smaller value of $\varepsilon_\Omega$ and the usefulness of the equality can be expanded to mesoscopic and even macroscopic systems. It should be borne is mind though that for $W > 0$ the expectation value shown in Eq. (1.1) is at most unity while the root –mean- square fluctuation around the expectation value as defined in Eq. (1.4) is always greater than unity.

The paper is organized as follows. In Sec II we review the nature of the thermal fluctuations in a ferromagnet . The moments of the work fluctuations are considered in Sec III and the resulting probability distribution in Sec IV. Our results are summarized in Sec V. Details about the average work and the shape of the work probability distribution are further elucidated in an Appendix.

**II. THERMAL FLUCTUATIONS IN A FERROMAGNET**

Recalling the statistical mechanics of a three- dimensional ferromagnet in an external magnetic field in the z- direction at very low temperatures , we note that the total magnetization $\vec{M}$ can be considered to be of fixed magnitude with most of its contribution coming from the large value $M_0$ in the z-direction. The transverse magnetization components ( the Goldstone modes of the system) constitute a two- dimensional vector $\vec{m}$ with components $m_1, m_2$ which at these low temperatures can be considered to be small fluctuations. The components of the magnetization vector satisfy the constraint

$$m_1^2 + m_2^2 + M_0^2 = M^2 \qquad (2.1)$$

with $M$ treated as a constant as explained. The free energy of the system ( non-linear sigma model) can be written as [24-26]

$$F = \frac{1}{V}\int d^3r \left[ \frac{J}{2}\left((\vec{\nabla}m_1)^2 + (\vec{\nabla}m_2)^2 + (\vec{\nabla}M_0)^2\right) - hM_0 \right] \qquad (2.2)$$

where $J$ is the strength of the coupling between the neighbouring transverse fluctuations and $V = L_\perp^2 L$ is the volume of the system. From Eq. (2.1), $M_0 \cong M - (m_1^2 + m_2^2)/2M$ and, hence, to quadratic order the free energy becomes

$$F = \frac{1}{V}\int d^3r \sum_{i=1}^{2}\left[ \frac{J}{2}(\vec{\nabla}m_i)^2 + \frac{h}{2M} m_i^2 \right] \qquad (2.3)$$

An important question is will the quadratic expression suffice for us . The higher- order terms are essential [26 ] when one is working near the critical point . In this case , we are deep in the ordered phase and the correlation length exponent $\nu$ can be taken to have the Gaussian value of ½. With this value of $\nu$, the higher- order terms will turn out to be irrelevant as seen for disordered electrons [ 27,28,30 ]. Consequently, we work here with the quadratic approximation for the free energy.

The correlation function $\langle m_i(\vec{x})m_j(\vec{x}+\vec{r}) \rangle$ defines the properties of the system and can be found from the corresponding correlation function in Fourier space. The dependence of the latter on the wave vector $\vec{k}$ can be easily deduced from Eq. (2.3) :

$$\langle m_i(\vec{k})m_j(\vec{k}') \rangle = \frac{k_B T V}{Jk^2 + \frac{h}{M}} \delta(\vec{k}+\vec{k}')\delta_{ij} \qquad (2.4)$$

where $T$ is the temperature at which the system is kept. The crossover that we mentioned earlier is evident from the above formula. For $h \to 0$ ( infinitesimal magnetic field), the correlation function is proportional to $k^{-2}$, which in co-ordinate space leads to a long-ranged $r^{-1}$ behaviour. For finite values of the magnetic field $h$, the correlation function falls off exponentially in space with a correlation length equal to $(MJ/h)^{1/2}$.

The system is now taken to be of finite extent in the z-direction by considering the extension in that direction to be of length $L$ which forces the magnetization fluctuations to be described by a Fourier series in the z- direction while the transform can still be used in the horizontal plane with area $L_\perp^2$. The dimensionless area is $A = L_\perp^2 / L^2$. In the z-direction large and small lengths are defined with respect to the correlation length . The Fourier decomposition of the magnetization fluctuation ( $m_1$ or $m_2$) which we will simply denote by $m(\vec{r})$, unless it is essential to differentiate between the components, can be written as

$$m(\vec{r}) = \int \frac{d^2 p}{(2\pi)^2} \sum_{n=1}^{\infty} \frac{2}{L} e^{i\vec{p}\cdot\vec{r}_\perp} \sin\frac{n\pi z}{L} m(\vec{p}, n) \qquad (2.5)$$

where $\vec{p}$ is a two- dimensional vector in the horizontal ($x-y$) plane and we have taken the magnetization fluctuations to vanish at the surfaces $z=0$ and $z=L$. The correlation function of Eq. (2.4) now becomes

$$\langle m(\vec{p},n) m(\vec{q},n') \rangle = \frac{k_B T V L}{J(p^2 + \frac{n^2\pi^2}{L^2}) + \frac{h}{M}} \delta(\vec{p}+\vec{q}) \delta_{nn'} \qquad (2.6)$$

Turning to the work done as the magnetic field is changed from an initial $h=0$ to a final value '$h$', we define

$$W_E = -\int_0^h dh' M_0 \cong -Mh + \int_0^h dh'(m_1^2 + m_2^2)/2 \qquad (2.7)$$

We need to comment on the definition of the work $W$ above for the magnetic system. In analogy with the work done for the expanding gas discussed in section I, one would have expected the differential work for the magnetic system to be $dW = hdM$. However, it is simpler to do calculations with Eq. (2.7). To this end we define a Gibbs free-energy $G = U - TS - hM$, so that for an isothermal process, the dissipative entropy satisfies $TdS - dQ = dW_E - dG$. Here $dW_E$ is an 'enthalpy' work, $U$ is the internal energy and $dQ$ is the heat energy. The Jarzynski equality (Eq. (1.1) ), when the magnetic work is $W_E$ of Eq. (2.7), becomes $\langle e^{-W_E/kT} \rangle = e^{-\Delta G/kT}$. The Gibbs free energy appears on the right- hand side of Eq. (1.1) when the work done is the 'enthalpy' work defined above. In what follows we drop the subscript $E$. Returning to Eq. (2.7), the first term is a constant and the second is the work fluctuation $W_{fl}$ to the lowest order. We are ignoring the higher- order terms arising from the binomial expansion of $(M^2 - m_1^2 - m_2^2)^{1/2}$ in Eq. (2.7). The average value of the fluctuations, $\langle W_{fl} \rangle$ is much smaller than $Mh$. If we calculate explicitly the contribution to $\langle W_{fl} \rangle$ coming from the next term to the one shown in Eq. (2.7), then it is seen that its contribution is smaller by $O(\langle W_{fl} \rangle / Mh)$, which, justifies the truncation shown in Eq. (2.7).

Our interest is in the probability distribution for $W_{fl} = \int_0^h dh'(m_1(\vec{r})^2 + m_2(\vec{r})^2)/2$, which we will access by calculating the various moments that are obtained by averaging over the magnetization fluctuations at a temperature $T$ and averaging over space. The $n$-th moment will be defined as

$$\langle W_{fl}^n \rangle = \left(\frac{1}{2MV}\right)^n \int d^3 r_1 \ldots \int d^3 r_n \langle (m_1^2(r_1) + m_2^2(r_1)) \ldots (m_1^2(r_n) + m_2^2(r_n)) \rangle \quad (2.8)$$

Since fluctuations in directions '1' and '2' are not correlated, the above can be written as

$$\langle W_{fl}^n \rangle = 2 \left(\frac{1}{2MV}\right)^n \int d^3 r_1 \ldots d^3 r_n \langle m^2(\vec{r}_1) m^2(\vec{r}_2) \ldots m^2(\vec{r}_n) \rangle \quad (2.9)$$

In the next section we shall calculate the moments ( actually only the cumulants ) and in Section IV we shall obtain the asymptotic form of the probability distribution.

### III. MOMENTS OF THE WORK FLUCTUATIONS

We start with the first moment which is the average of the work fluctuations ($n$=1 in Eq. (2.9)) and find

$$\langle W_{fl} \rangle = \frac{1}{MV} \int_0^h dh' \int d^3 r \langle m(\vec{r})^2 \rangle$$

$$= \frac{1}{MV} \int_0^h dh' \int d^3 r \sum_{n_1, n_2} \frac{4}{L^2} \sin\frac{n_1 \pi z}{L} \sin\frac{n_2 \pi z}{L} \int \frac{d^2 p}{(2\pi)^2} \int \frac{d^2 q}{(2\pi)^2} e^{i(\vec{p}+\vec{q}) \cdot \vec{r}} \langle m(\vec{p}, n_1) m(\vec{q}, n_2) \rangle$$

$$= \frac{2}{MVL^2} \int_0^h dh' \int d^3 r \sum_{n=1}^\infty \int \frac{d^2 p}{(2\pi)^2} \sin^2\frac{n\pi z}{L} \frac{k_B TVL}{J(p^2 + \frac{n^2 \pi^2}{L^2}) + \frac{h'}{M}}$$

$$= \frac{L_\perp^2}{L^2} \frac{L^2}{MJ} \int_0^h dh' \sum_{n=1}^\infty \int \frac{d^2(pL)}{(2\pi)^2} \frac{k_B T}{p^2 L^2 + n^2 \pi^2 + \frac{h'L^2}{MJ}} \quad (3.1)$$

From now on we label the dimensionless wave number $pL$ as $q$ and define

$$\gamma^2 = p^2 L^2 + \frac{h'L^2}{MJ} = q^2 + \frac{h'L^2}{MJ}, \qquad (3.2)$$

After performing the summation over '$n$' and introducing an ultraviolet cut-off at $p = \Lambda$, we write Eq. (3.1) as

$$\langle W_{fl} \rangle = k_B T \frac{L_\perp^2}{L^2} \frac{L^2}{MJ} \int_0^h dh' \int_0^{\Lambda^2 L^2} \frac{d(q^2)}{4\pi} \frac{1}{2\gamma}(\coth\gamma - \frac{1}{\gamma})$$

$$= \frac{1}{4\pi} k_B T A L_h^2 \int_0^1 dx \int_{\left(\frac{hL^2}{MJ}\right)^{1/2}}^{\Lambda L} d\gamma (\coth\gamma - \frac{1}{\gamma}) \qquad (3.3)$$

where as defined before $A = L_\perp^2 / L^2$ and $L_h^2 = hL^2/MJ$. Evaluating the integral in Eq. (3.3), we now have

$$\langle W_{fl} \rangle = k_B T \frac{A L_h^2}{4\pi} \int_0^1 dx \left[ \ln\frac{\sinh\Lambda L}{\Lambda L} - \ln\frac{\sinh\sqrt{xL_h^2}}{\sqrt{xL_h^2}} \right]$$

$$= k_B T \frac{A L_h^2}{4\pi} \left[ \Lambda L - \ln(2\Lambda L) - \int_0^1 dx \ln\frac{\sinh\sqrt{xL_h^2}}{\sqrt{xL_h^2}} \right] \qquad (3.4)$$

The first two terms in the square bracket above, which are cut-off dependent, arise from the divergence of the integral. We note that this divergence stems from the large wave- number dependence of the integrand and thus corresponds to the short- ranged correlations. It should be noted that the average work is proportional to the volume of the system because of this divergence and is very different from the higher moments. We will return to this issue in the Appendix. The crossover effect is present in the finite part represented by the integral which is zero for $L_h^2 \ll 1$, while for $L_h^2 \gg 1$ the dominant term comes from the growing exponential inside the log and it is easily seen to be $2L_h / 3$. We note that the high-momentum divergence that we see in the first moment will not occur in the higher moments. It should also be noted that in the high- field limit the behaviour $L_h^3$ that has emerged will remain throughout regardless of the degree of the moment.

We now turn to the second moment ( actually the cumulant ) which will be the prototype for all other cumulants since it will not have the divergence that plagues the first moment. We will do this in some detail and then generalize to the higher moments. Setting $n = 2$ in Eq. (2.9), we consider

$$\langle W_{\text{fl}}^2 \rangle = 2\left(\frac{1}{2MV}\right)^2 \int_0^h dh' \int_0^h dh'' \int d^3r_1 \int d^3r_2 \, \langle m_{h'}^2(r_1) m_{h''}^2(r_2) \rangle \qquad (3.5)$$

The subscripts associated with the two different $m^2(r)$ above make it clear that the process $h'=0$ to $h'=h$ will in general be different for the different $m^2(\vec{r})$. A Gaussian distribution for the magnetization fluctuations will imply a factorization of the correlation function shown above and we have for the second-order cumulant

$$\langle W_{\text{fl}}^2 \rangle_{\text{cum}} = 2 \times 2\left(\frac{1}{2MV}\right)^2 \int_0^h dh'' \int_0^h dh' \int d^3r_1 \int d^3r_2 \, \langle m_{h'}(r_1) m_{h''}(r_2) \rangle^2 \qquad (3.6)$$

In the above we have shown the first factor of 2 separately as it is a combinatorial factor that will change with $n$. To find the equilibrium equal-time correlation that we need in Eq. (3.6), it is simplest to do the calculation in Fourier space and evaluate $\langle m_h(\vec{k},t) m_{h'}(-\vec{k},t) \rangle$.

We begin by writing the evolution equation for the Fourier transform $m_h(\vec{k},t)$. The dynamics is the usual Langevin dynamics with the noise having the appropriate statistical property to give the equilibrium correlation function of Eq. (2.4). Explicitly,

$$\dot{m}_h(\vec{k},t) = -\Gamma(Jk^2 + \frac{h}{M}) m(\vec{k},t) + \varsigma(\vec{k},t) \qquad (3.7)$$

where $\varsigma(\vec{k},t)$ is the delta-correlated Gaussian white noise with the correlation function

$$\langle \varsigma(\vec{k}_1,t_1) \varsigma(\vec{k}_2,t_2) \rangle = 2\Gamma k_B TV (Jk_1^2 + \frac{h}{M}) \delta(\vec{k}_1 + \vec{k}_2) \delta(t_1 - t_2) \qquad (3.8)$$

After integrating,

$$m_h(\vec{k},t) = \int dt' \varsigma(\vec{k},t') e^{-\Gamma(Jk^2 + \frac{h}{M})(t-t')} + m_h(\vec{k},0) e^{-\Gamma(Jk^2 + \frac{h}{M})t} \qquad (3.9)$$

The equilibrium correlation function is obtained as the equal-time correlation function at long times and is seen to be

$$\langle m_h(\vec{k}) m_{h'}(\vec{k}') \rangle = \frac{k_B TV}{Jk^2 + \frac{h+h'}{2M}} \delta(\vec{k} + \vec{k}'), \qquad (3.10)$$

giving the standard equilibrium correlation function of Eq. (2.4) when $h = h'$.

Rreturning to Eq. (3.6), we write

$$\langle m_{h'}(\vec{r}_1)m_{h''}(\vec{r}_2)\rangle = \sum_{n_1=1}^{\infty} \frac{4}{L^2}\sin\frac{n_1\pi z}{L}\sum_{n_2=1}^{\infty}\sin\frac{n_2\pi z}{L}\int\frac{d^2p}{(2\pi)^2}\int\frac{d^2q}{(2\pi)^2}e^{i\vec{p}\cdot\vec{r}_{1\perp}+i\vec{q}\cdot\vec{r}_{2\perp}}\langle m_{h'}(\vec{p},n_1)m_{h''}(\vec{q},n_2)\rangle$$

$$= \sum_{n=1}^{\infty}\frac{2}{L}\sin\frac{n\pi z_1}{L}\sin\frac{n\pi z_2}{L}\int\frac{d^2p}{(2\pi)^2}e^{i\vec{p}\cdot(\vec{r}_{1\perp}-\vec{r}_{2\perp})}\frac{k_BTV}{J(p^2+\frac{n^2\pi^2}{L^2})+\frac{h''+h'}{2M}} \quad (3.11)$$

To evaluate the integral in Eq. (3.6), we need another such factor with $n$ replaced by $n'$ and $\vec{p}$ replaced by $\vec{q}$. The integral over $z_1$ will then yield a factor of $L\delta_{nn'}/2$ and the integration over $z_2$ gives $L/2$. Similarly, integration over $r_{1\perp}$ yields $\delta(\vec{p}+\vec{q})$ and the integration over $r_{2\perp}$ gives $L_\perp^2$. This allows us to write Eq. (3.6) as

$$\langle W_{fl}^2\rangle_{cum} = 2\times 2\left(\frac{1}{2MV}\right)^2\int_0^h dh''\int_0^h dh'\sum_{n=1}^{\infty}\int\frac{d^2p}{(2\pi)^2}\left[\frac{k_BTV}{J(p^2+\frac{n^2\pi^2}{L^2})+\frac{h''+h'}{2M}}\right]^2 \quad (3.12)$$

Carrying out the rescalings described in the calculation of $\langle W_{fl}\rangle$ above, we write Eq. (3.12) as

$$\langle W_{fl}^2\rangle_{cum} = 2\times\frac{1}{2}(2k_BTL_h^2)^2 A\int_0^1 dx''\int_0^1 dx'\sum_{n=1}^{\infty}\int\frac{d(q^2)}{4\pi}\left(\frac{1}{q^2+n^2\pi^2+L_h^2\frac{x''+x'}{2}}\right)^2$$

$$= \frac{1}{4\pi}(2k_BTL_h^2)^2 A\int_0^1 dx''\int_0^1 dx'\sum_{n=1}^{\infty}\frac{1}{n^2\pi^2+L_h^2\frac{x''+x'}{2}}$$

$$= \frac{1}{4\pi}(2k_BTL_h^2)^2 A\int_0^1 dx''\int_0^1 dx'\frac{1}{2\gamma'}(\coth\gamma'-\frac{1}{\gamma'}) \quad (3.13)$$

The $\gamma'$ in the last line of Eq. (3.13) is defined as $\gamma'^2 = L_h^2(x''+x')/2$. We now examine this formula in the two limiting cases of $L_h^2 \ll 1$ and $L_h^2 \gg 1$.

    i)    $L_h^2 \ll 1$: In this limit $\gamma' \to 0$ and the first non-vanishing term of the integrand in Eq. (3.13) is seen to be 1/3 and, hence,

$$\langle W_{fl}^2\rangle_{cum} = \frac{1}{4\pi}(2k_BTL_h^2)^2 A\left[\frac{1}{6}+O(L_h^4)\right] \quad (3.14)$$

ii)     $L_h^2 \gg 1$: In this limit $\gamma' \to \infty$ and the leading contribution comes from the region where $\coth \gamma' = 1$ and we find

$$\langle W_{fl}^2 \rangle_{cum} = \frac{1}{4\pi}(2k_B T)^2 A L_h^3 \int_0^1 dx'' \int_0^1 dx' \frac{1}{2(x''+x')^{1/2}} = \frac{1}{4\pi}(2k_B T)^2 A L_h^3 \frac{4}{3}(\sqrt{2}-1) \quad (3.15)$$

In this limit the next term is of $O(L_h^2)$.

We note that while the finite part of $\langle W_{fl} \rangle$ would be $O(L_h^2)$ [ it is accidental that the coefficient is zero] in the small $L_h$ limit, the corresponding dependence of $\langle W_{fl}^2 \rangle_{cum}$ is $O(L_h^4)$ (Eq. (3.14)) and for $\langle W_{fl}^n \rangle_{cum}$ it will be $O(L_h^{2n})$ as we will argue below. However, for very large values of $L_h$, the dependence of $\langle W_{fl}^n \rangle_{cum}$ will always be $L_h^3$ regardless of $n$. For any '$n$', we can write

$$\langle W_{fl}^n \rangle_{cum} = 2^{n-1}(n-1)! \times 2\left(\frac{1}{2MV}\right)^n \int_0^h dh_1 ... \int_0^h dh_n \int d^3 r_1 ... \int d^3 r_n \langle m_{h_1}(\vec{r}_1) m_{h_2}(\vec{r}_2) \rangle ...... \langle m_{h_n}(\vec{r}_n) m_{h_1}(\vec{r}_1) \rangle$$
(3.16)

Inserting the appropriate Fourier expansions and performing the spatial integrations, we obtain

$$\langle W_{fl}^n \rangle_{cum} = \frac{1}{4\pi}(n-1)! A (2k_B T L_h^2)^n \int_0^1 dx_1 ... \int_0^1 dx_n \sum_{s=1}^{\infty} \int_0^{\infty} d(q^2) \frac{1}{q^2 + s^2 \pi^2 + L_h^2 \frac{x_1+x_2}{2}} \frac{1}{q^2 + s^2 \pi^2 + L_h^2 \frac{x_2+x_3}{2}} .... $$

$$..... \frac{1}{q^2 + s^2 \pi^2 + L_h^2 \frac{x_n+x_1}{2}} \quad (3.17)$$

In the limit of infinitesimal fields, $L_h \to 0$, we find ( $n \geq 2$)

$$\langle W_{fl}^n \rangle_{cum} (L_h \to 0) = \frac{\pi}{4}(n-2)! A \left(\frac{2k_B T}{\pi^2}\right)^n L_h^{2n} \varsigma(2n-2) \quad (3.18)$$

On the other hand, in the limit of large $L_h$, the scaling behaviour can be inferred by noting that the integrand in Eq. (3.17) can be written as partial fractions where each term has the form $(q^2 + s^2 \pi^2 + L_h^2 \frac{x_i+x_j}{2})^{-1}$ multiplied by $n-1$ terms of the form $(L_h^2 \frac{x_l-x_k}{2})^{-1}$. These $n$-1 terms reduce the power of $L_h^2$ in the prefactor of Eq. (3.17) to unity and the remaining $L_h$ dependence comes from the sum and integral $\sum_{s=1}^{\infty} \int d(q^2)[q^2 + s^2\pi^2 + L_h^2 \frac{x_i + x_j}{2}]^{-1}$, which after summation over '$n$' can be written in the leading order for large $L_h^2$ as $\int d(q^2)/[q^2 + L_h^2 \frac{x_i + x_j}{2}]^{1/2}$. This integral scales as $L_h$ and hence for $L_h \gg 1$, the leading term of Eq. (3.17) is

$$\left\langle W_{fl}^n \right\rangle_{cum} (L_h \to \infty) = (n-2)! A \left( \frac{2k_B T}{\pi^2} \right)^n C_n L_h^3 \qquad (3.19)$$

where $C_n$ is a number of order unity. The dependence on $h$ follows from the general result that the Goldstone-mode-dominated susceptibility scales as $L_h^{-1}$ [29,30]. We note that all the moments are proportional to the system size $L_\perp^2 L$ as happens always in a system with short-ranged correlations. In sharp contrast, when one is dealing with long-ranged correlations ($L_h \to 0$), all the moments (except the average work) are proportional to $L_\perp^2 / L^2$ as shown in Eq. (3.18).

What kind of a crossover function are we finding? To answer this it is easiest to examine higher-order terms in Eq. (3.17) as $L_h$ is increased. Expanding the denominator of each factor in the integrand, it is clear that the series is going to be alternating and the first correction is

$$\left\langle W_{fl}^n \right\rangle_{cum} = \left\langle W_{fl}^n \right\rangle_{cum} (L_h \to 0) \left[ 1 - \frac{n-1}{2} \frac{\zeta(2n-4)}{\zeta(2n-2)} L_h^2 + O(L_h^4) \right] \qquad (3.20)$$

An interpolation which guarantees that coefficients come out correctly for large $n$ (in practice $n > 4$, when the zeta functions are effectively unity) is

$$\left\langle W_{fl}^n \right\rangle_{cum} = \left\langle W_{fl}^n \right\rangle_{cum} (L_h \to 0) \left[ 1 + L_h^2 \right]^{3/2} / \left[ 1 + \frac{L_h^2}{2} \right]^n \qquad (3.21)$$

For the special case of $n=2$, we introduce the crossover function $f(L_h)$ which gets the coefficients correctly in both the small and large $L_h$ limits and write

$$\left\langle W_{fl}^2 \right\rangle_{cum} = (k_B T)^2 A f(L_h) \qquad (3.22)$$

An explicit interpolation form for $f(L_h)$ is

$$f(L_h) = \frac{L_h^4}{6\pi\left(1 + \frac{L_h}{8(\sqrt{2}-1)}\right)} \quad (3.23)$$

which as we will see below is the function that will describe the crossover of the fluctuation $\varepsilon_\Omega$, introduced in Eq. (1.6). Having obtained the moments we now construct the probability distribution in the next section.

## IV THE PROBABILITY DISTRIBUTION

We find the work probability distribution $\rho(W)$ by constructing a cumulant generating function $K(t)$ defined by

$$K(t) = \sum_{n=1}^{\infty} \frac{\langle W_{fl}^n \rangle_{cum} \, t^n}{n!} \quad (4.1)$$

In this sum, the terms $n \geq 2$ are determined from Eqs. (3.18) and (3.21), while for $n=1$, we need the leading term ( proportional to the size of the system ) of Eq. (3.4). The probability distribution function is then obtained by taking the inverse Laplace transform

$$\rho(W) = \int_0^\infty e^{-Wt + K(t)} dt \quad (4.2)$$

The integral in Eq. (4.2) will be evaluated by the method of steepest descent, since the evaluations of $K(t)$ or $\rho(W)$ can not be done exactly. In the sum shown in Eq. (4.1), the first two terms determine the Gaussian approximation to the distribution function. The higher- order terms ($n > 2$) are responsible for the tails of the distribution function. To get the tails accurately, we need the larger values of $n$ and as a result $n-1$ can be replaced by $n$ and the zeta functions set to unity ( $\varsigma(4) = \pi^4/90$ is already very close to unity). For $W > 0$, we have

$$K(t) = \frac{\pi}{4} A(1 + L_h^2)^{3/2} \sum_n (k_B T)^n \left(\frac{2L_h^2}{\pi^2(1 + \frac{L_h^2}{2})}\right)^n \frac{t^n}{n^2}$$

$$= a\sum_n \frac{(bt)^n}{n^2} \quad (4.3)$$

with

$$a = \pi A(1+L_h^2)^{3/2}/4, b = k_B T \frac{4L_h^2}{\pi^2(2+L_h^2)} \qquad (4.4)$$

It should be noted that the $K(t)$ of Eq. (4.3) gives an average value of $W$ that is $\langle W \rangle = ab$, different from the result obtained in Eq. (3.4). This will be important when we discuss the Gaussian approximation below.

From Eq. (4.3), we see that

$$K'(t) \equiv \frac{dK}{dt} = \frac{a}{t}\ln\frac{1}{(1-bt)} \qquad (4.5)$$

The saddle-point method for obtaining $\rho(W)$ requires finding the saddle point $t_0$ where

$$W = K'(t_0) \qquad (4.6)$$

and writing

$$\rho(W) = Ce^{-Wt_0 + K(t_0)} \qquad (4.7)$$

where $C$ is a constant obtained from normalization. We need to find $t_0$ from the requirement

$$W = -\frac{ab}{bt_0}\ln(1-bt_0) = -\frac{ab}{x_0}\ln(1-x_0) \qquad (4.8)$$

where $x = bt$. The scale of $W$ is set by $ab$. We need to emphasize again that this scale is not necessarily the average value of $W_{fl}$. For small values of $L_h^2$ the dimensionless scale ($ab/k_B T$) is set by $AL_h^2/2\pi$, which scales as $hV/L^3$. For large values of $L_h^2$, the scale is set by $AL_h^3/\pi$ which scales as $h^{3/2}V$. This scale, like the average work in Eq. (3.4), is proportional to the system size. In this case, things are reasonably clear- all the action is centred around the scale '$ab$'. Regions close to this scale and far away from it can be explored by expanding the right-hand side of Eq. (4.8) in a Taylor series and determining $x_0$. The probability distribution follows from Eq. (4.7). We represent the scale factor by the crossover form (see Eq. (4.4))

$$\frac{ab}{k_B T} = \frac{A}{\pi}\frac{\left(1+L_h^2\right)^{3/2}}{\left(1+\frac{2}{L_h^2}\right)} = \frac{A}{\pi}S(L_h) \qquad (4.9)$$

Writing the dimensionless work in units of $k_B T$ in Eq. (4.8), we find

$$\pi W = -AS(L_h)\frac{\ln(1-x_0)}{x_0} \qquad (4.10)$$

For both $L_h \gg 1$ and $L_h \ll 1$, large values of $W$ correspond to $x_0 \cong 1$. Writing $x_0 = 1-\varepsilon$, we see that $\varepsilon \cong \exp[-\pi W / AS]$. The integral of Eq. (4.5) can be written as

$$K(t) = -a\left[\ln x \ln(1-x) + \int \frac{\ln x}{1-x} dx\right] + K_0 \qquad (4.11)$$

where $K_0$ is a constant. All terms in $K(1-\varepsilon)$ are at least $O(\varepsilon)$ and hence the asymptotic large $W$ behaviour of $\rho(W)$ is $\exp(-W/b)$. The scale of $W$ in Eq. (4.8) changes with changing $L_h$ but the form of $\rho(W)$ is unaltered.

For $x_0 \ll 1$, one is in the vicinity of $W = AS(L_h)/\pi$, which is proportional to $L_h^2$ for small magnetic fields and to $L_h^3$ for large fields. In this range, $x_0 \cong 2\left(\frac{W\pi}{AS(L_h)} - 1\right)$ and the distribution, in this approximation, is a Gaussian centered at $AS(L_h)$. This, as expected, is different from the exact Gaussian obtained by keeping the $n=1$ and $n=2$ terms in Eq. (4.1).

The limit of very small $W$ needs to be handled carefully and we note that the only way one can achieve small values of $-\frac{\ln(1-x_0)}{x_0}$ is by going to large negative values of $x_0$ so that Eq. (4.8) becomes

$$\pi W = AS(L_h)\frac{\ln(1+|x_0|)}{|x_0|} \qquad (4.12)$$

Inverting, we obtain

$$|x_0| \cong AS(L_h)\frac{\ln(AS(L_h)\pi^{-1}W^{-1})}{\pi W} \qquad (4.13)$$

Integration of Eq. (4.5) for large negative values of $t$, yields

$$K(t) = -\frac{a}{2}\left(\ln|x|\right)^2 \qquad (4.14)$$

For $W \to 0$, we use Eq. (4.13) to calculate the exponent $-Wt_0 + K(t_0)$ (we are not showing any prefactors since they will not be relevant) and get

$$\rho(W \to 0) \propto e^{-\frac{a}{2}\left(\ln\frac{AS(L_h)}{\pi W}\right)^2 + a\ln\left(\frac{AS(L_h)}{\pi W}\right)} \qquad (4.15)$$

For $W$ in the vicinity of $ab$, we have the Gaussian form (note that this Gaussian form stems from the approximate form of $K(t)$ in Eq. (4.3)) given by

$$\rho(W) \propto \exp\left(-\frac{(W-ab)^2}{ab^2}\right) \quad (4.16)$$

Using the $n=1$ and $n=2$ terms of Eq. (4.1), we obtain the more general Gaussian form,

$$\rho(W) \propto e^{-\frac{(W-\langle\langle W\rangle\rangle)^2}{2\langle W_{fl}^2\rangle}} = e^{-\frac{\langle W\rangle^2}{2\langle W_{fl}^2\rangle}\left(\frac{W}{\langle W\rangle}-1\right)^2} = e^{-\frac{\langle W\rangle^2}{2Af(L_h)}\left(\frac{W}{\langle W\rangle}-1\right)^2} \quad (4.16a)$$

For $W \gg ab$, we find

$$\rho(W) \propto \exp(-W/b) \quad (4.17)$$

Eqs. (4.15)-(4.17) constitute the basis for the primary result of this paper quoted in Eq. (1.6).

Specifically, for the work distribution in the case of long-ranged correlation ($L_h < 1$) we obtain:

i) $\quad W \ll AL_h^2/2\pi, \quad \rho(W) \propto \exp\left[-\frac{\pi A}{8}\left(\ln\frac{AL_h^2}{2\pi W}\right)^2 + \frac{\pi A}{4}\ln\left(\frac{AL_h^2}{2\pi W}\right)\right] \quad (4.18a)$

ii) $\quad W \approx \langle W \rangle = AL_h^2/2\pi, \quad \rho(W) \propto \exp\left[-\frac{\pi A\left(\frac{2\pi W}{AL_h^2}-1\right)^2}{4}\right] \quad (4.18b)$

iii) $\quad W \gg AL_h^2/2\pi \quad \rho(W) \propto \exp\left(-\pi^2 W/2L_h^2\right) \quad (4.18c)$

The results i) and iii) are identical to what was reported in Kirkpatrick *et al*. [23]. For ii), the Gaussian form shown in Eq. (4.18b) is based on Eq. (4.16) while the form shown in reference [23] was based on Eq. (4.16a). The difference does not matter in establishing the primary point, which is the scale of the probability distribution. The difference between Eqs. (4.16a) and (4.18b) will be commented upon in the Appendix.

The corresponding results $\rho_{SR}$ for the short-ranged correlations, as follows from Crooks and Jarzynski [12], are

$$\rho_{SR}(W \to 0) \propto e^{-\frac{3N}{2}\ln\left(\frac{\langle W\rangle}{W}\right)} \quad (4.19a)$$

$$\rho_{SR}(W \approx \langle W\rangle) \propto e^{-\frac{3N}{2}\left(\frac{W}{\langle W\rangle}-1\right)^2} \quad (4.19b)$$

$$\rho_{SR}(W >> <W>) \propto e^{-\frac{3NW}{2<W>}} \qquad (4.19c)$$

The important difference between the above short- ranged probability distributions and the long- ranged ones of Eqs. (4.18a) and (4.18b) lies in the scale of the distribution. In the long-ranged case it is set by $A$ which is proportional to $V/L^3$ and for the short- ranged case it is set by $N$ which is equal to $V/a^3$ where $a$ is the lattice spacing. For $W << \langle W \rangle$, the distributions are similar. In the evaluation of the moments of $\Omega$, it is the $W \leq <W>$ parts which matter and hence the suppression of the probability distribution for the short- ranged correlation leads to a larger $\varepsilon_\Omega$. If we consider a specific case where $L_\perp = 10a$ (typical protein) and $L = 4a$ which makes $A = 6.25$ and two values of $W$ such that $W << \langle W \rangle$ and $W \approx \langle W \rangle$, we should be able to make our point. The value of $N$ for this case can be found from $Na^3 = V = L_\perp^2 L$ and is seen to be 400. For $W = \langle W \rangle / 100$, $\rho(W)/\rho_{SR}(W) \approx e^{1317}$ and for $W = \langle W \rangle / 2$, this ratio is $e^{149}$. This makes the point that the systems with long-ranged correlations have a much broader probability distribution for the work done.

We now turn to the calculation of the fluctuation $\varepsilon_\Omega$ defined in Eq. (1.5). We use the distribution shown in Eq. (4.16a) for this purpose. Carrying out the Gaussian integrals,

$$\langle e^{-nW} \rangle = e^{n^2 A f(L_h)/2} \qquad (4.20)$$

Consequently,

$$\varepsilon_\Omega = \frac{\langle \Omega^2 \rangle}{\langle \Omega \rangle^2} - 1 \cong \frac{\langle \Omega^2 \rangle}{\langle \Omega \rangle^2} = e^{A f(L_h)} \qquad (4.21)$$

with $f(x)$ defined in Eqs. (3.22) and (3.23). We have thus established Eq. (1.6) which was stated to be the primary result of the paper. If we take the example cited above ($A = 6.25$, $N = 400$) then for the long-ranged case ($L_h << 1$), $\varepsilon_\Omega \approx e^{2L_h^4/3\pi}$ which can be quite close to unity for $L_h << 1$. For the short –ranged case, using Eq. (1.5) with $N = 400$ and assuming small $\alpha$, we find $e^{600\alpha^2}$ which is astronomical unless $\alpha$ is very small. The root-mean-square fluctuation about the average of $\Omega$ is $e^{300\alpha^2}$ much greater than $<\Omega>$ itself which is $e^{-600\alpha^2}$. If one reduces the number of particles still further (in terms of system size, considering nano scales and beyond) one can make the root-mean- square fluctuation smaller for short-ranged fluctuations but it will still be greater than the average value.

It is necessary to ask the question how small must the external magnetic field be to allow us to use the long-ranged approximation, $L_h^2 << 1$. For this we need an estimate for $L$, which can be taken to be the typical size of ferromagnetic domains. Free- energy considerations restrict the domain size to $10^{-7}$ m to $10^{-8}$ m. To be definite, we take $10^{-8}$ m as our $L$. If we denote a microscopic length by $l$, then $L = 100l$. Writing $L_h^2 = hL^2 / MJ = L^2 / L_H^2$, we have $L_H^2 = MJ / h = M^2J / Mh = k_B T l^2 / \mu_B h$, where in the last step we have replaced the magnetization in the denominator by the Bohr magneton $\mu_B$ and estimated $JM^2$ by appealing to Eq. (2.3) and writing $k_B T$ for the free energy. At $T = 100$ K, $k_B T \approx 10^{-21}$ Joules and with $h$ in teslas, $\mu_B \approx 10^{-23}$ Joules/tesla. We thus estimate $L_H^2 \approx 100 l^2 / h$ and $L_h^2 \approx 100h$, so that the condition $L_h^2 < 1$ implies $h < 10^{-2}$ teslas which is $10^2$ Gauss.

It should be noted that the kind of experiments used to test the Jarzynski equality have been primarily single-molecule experiments (*e.g.*, Refs. [31-33]) where the probability distribution has been used to determine the free-energy difference between initial and final states [*e.g.*, the folded and unfolded states of a DNA hairpin] or the energy fluctuations in a single harmonic oscillator driven out of equilibrium by an external torque. An example of the latter is the thermal rheometer [34-36] which is a torsion pendulum whose minute angular displacements are measured by a highly sensitive interferometer. The pendulum is driven out of equilibrium by an external torque which amounts to a few pico newton-metres. The thermal fluctuations amount to a root-mean-square angular displacement of a few nano radians. This gives an idea of the scale of the system. The torsion pendulum operates in the linear regime, the probability distributions are Gaussian and the test of the fluctuations dissipation theorems depends only on the width of the distribution. What we are pointing out is that even for a mesoscopic system (the example given above) the Jarzynski equality will be difficult to test because of the large fluctuations, but if the fluctuations are long-ranged then even for a macroscopically large system ($N \square 10^{23}$), the fluctuation $\varepsilon_\Omega$ can be reduced to be of $O(1)$.

The result for $\varepsilon_\Omega$ shown in Eq. (4.21) is not a consequence of the Gaussian form for the probability distribution. We will get a similar result if we use the low-$W$ form of the distribution Eq. (4.18a). We are interested in calculating $\langle \Omega^n \rangle$ and we will use the probability distribution of Eq. (4.18a) to the leading order only. We need

$$\langle \Omega^n \rangle = \int \rho(W) \Omega^n dW / \int \rho(W) dW$$

$$= \int \rho(W) e^{-nW} dW / \int \rho(W) dW \qquad (4.22)$$

The calculation will entail ignoring all non-exponential pre-factors and hence the normalizing integral in the denominator will contribute unity. The numerator will be calculated in the saddle-point approximation described before. We need the integral

$$I_n = \int_0^\infty \exp\{-\frac{\pi A}{8}\left[\left(\ln\frac{AL_h^2}{2\pi W}\right)^2 + \frac{8nW}{\pi A}\right]\}dW = \int_0^\infty \exp\left[-\frac{\pi A}{8}g_n(W)\right]dW \quad (4.23)$$

In terms of the variable $x = 2\pi W / AL_h^2$, the function $g(W)$ can be written as

$$g_n(x) = \left(\ln\frac{1}{x}\right)^2 + n\frac{4L_h^2}{\pi^2}x = \left(\ln\frac{1}{x}\right)^2 + n\mu x \quad (4.24)$$

where $\mu = 4L_h^2 / \pi^2 \ll 1$ in the infinitesimal magnetic-field range where $L_h^2 \ll 1$. For very small values of $\mu$, the above function has a minimum very near $x = 1$ and this dominates the integrand in Eq. (4.24). The minimum is at $x_m = 1 - \delta$, where to the lowest order in $\mu$, it is seen that $\delta = \mu/2$. Up to $O(\mu^2)$, the minimum value of $g_n(x)$ is found as

$$g_n(x_m) = n\mu - \frac{n^2\mu^2}{4} \quad (4.25)$$

Keeping only the exponential terms, we have in this approximation

$$I_n \cong \exp\left[-\frac{\pi A}{8}(n\mu - \frac{n^2\mu^2}{4})\right] \quad (4.26)$$

Thus we have,

$$\varepsilon_\Omega = \frac{\langle\Omega^2\rangle}{\langle\Omega\rangle^2} - 1 = \frac{I_2}{I_1^2} - 1 \cong \exp\left[\frac{\pi A}{16}\mu^2\right] = \exp\left[A\frac{L_h^4}{\pi^3}\right] \quad (4.27)$$

We have obtained for $L_h \ll 1$, the same result as before with a different numerical factor.

Turning to the other extreme, where $L_h^2 \gg 1$ and the energy scale as well as the scale of the distribution for small $W$ is set by the system size, we note that

i)     For $W \ll AL_h^3$:     $\rho(W) \propto \exp-\frac{\pi AL_h^3}{8}\left[\ln\frac{AL_h^3}{\pi W}\right]^2$     (4.28a)

ii)     For $W \cong AL_h^3$     $\rho(W) \cong \exp\left[-\frac{(W - 4\pi AL_h^3)^2}{16\pi AL_h^3}\right]$     (4.28b)

iii)     For $W \gg AL_h^3$     $\rho(W) \cong \exp(-\pi^2 W / 4)$     (4.28c)

In the range of large magnetic field (short-ranged correlations) the factor $AL_h^3$ is like $h^{3/2}V$ and the probability distribution is tremendously suppressed for the relevant regions. This completes what we wanted to describe- the passage of the work probability distribution from a broad distribution for long ranged correlations to a very sharp distribution around a mean value for short ranged correlations. The passage can be experimentally checked by tuning the external magnetic field for a ferromagnet at temperatures well below the Curie temperature.

## V. CONCLUSIONS

The probability distribution for the work done in taking a system from one thermodynamic state to another has often been studied analytically [7,12,14], but surprisingly the fluctuations around the average of the quantity $\Omega = \exp(-W/k_B T)$ has not been looked at before the investigation in Ref. [13] . There was general consensus that the probability distribution would be sharply centred around some average value $\bar{W}$ which would be of $O(N)$ where $N$ is the system size and would have a exponential ( $\exp(-W/k_B T)$ ) tail for $W \gg \bar{W}$ indicative of dominance of rare events. The fact that a quantity like $\Omega$, which is vanishingly small where the probability distribution peaks, can have very large fluctuations was generally overlooked. In Ref. [13], it was pointed out that this large fluctuation, as would follow from the explicit distributions in Refs. [12] and [14], is primarily a feature of the short-range correlations of the fluctuations. It was also pointed out that if the correlations were long ranged , as happens in NESS for driven fluids , then results would be substantially different. What was clarified in Ref. [23] was the fact that long- ranged correlations , whether coming from NESS in driven fluids or from Goldstone modes in ferromagnets at low temperatures, have the ability to set the scale of the work probability distribution to much lower values and thus provide a much broader distribution. For $W \to 0$, the distribution $\rho(W) \to 0$ as $\rho(W) \propto \exp[-V(\ln\frac{\langle W \rangle}{W})^n / L^3]$, where $n$ is a number which is system specific. The scale is reduced from $V$ to a potentially much smaller quantity $V/L^3$ where $L$ can be several lattice spacings and thus for small systems at least, the fluctuations in $\Omega$ are significantly reduced, and when $L \square L_\perp$ it is very strongly reduced even for macroscopic systems. It would be interesting to revisit earlier experiments [31-36] which confirmed or utilized the Jarzynski equality, Eq. (1.1), with the large fluctuations in mind .

Ferromagnets have the interesting feature that by tuning the external magnetic field from infinitesimal to finite values the correlations can be changed from long- ranged ones to short- ranged ones. Consequently, it would be the ideal system for studying the crossover in the fluctuations around the mean value that appears on one side of the Jarzynski equality [4]. We have provided the explicit forms of the probability distributions at small fields ( long- ranged correlations ) which means fields less than 100 Gauss in practical terms in Eqs. ( 4.18a) to (4.18c) and at fields which are larger than 100 Gauss ( short ranged correlations ) in Eqs. (4.28a) to (4.28c). The fluctuation is seen to be $O(\exp(\frac{L_\perp^2}{L^2} L_h^4))$ for small fields ( small $L_h^2$ ). For larger magnetic fields, when the correlations are short ranged , this small $W$ behaviour is severely suppressed and the fluctuation $\varepsilon_\Omega$ is enormous and $O(\exp(h^{3/2} L_\perp^2 L))$ as happens in other short ranged systems. The ferromagnet at low temperatures provide a unique opportunity to study the crossover shown in Eq. (4.21) by tuning the external magnetic field.

**ACKNOWLEDGMENTS**:   Discussions with Bob Dorfman are gratefully acknowledged. This work is supported by the National Science Foundation under Grant number DMR-1401449. One of the authors (JKB) is grateful to the Burgers Program for Fluid Dynamics at the University of Maryland and an APS-IUSSTF professorship award which made the collaboration possible.

**APPENDIX:**    AVERAGE WORK AND SHAPE OF THE PROBABILITY DISTRIBUTION

We want to discuss the consequences of the fact that the average, $<W>$, of the work done by the magnetization fluctuations is of $O(V)$ , as seen from Eq. (3.4), even when the correlations are long-ranged making all the other moments of the work fluctuations proportional to $L_\perp^2 / L^2 = V / L^3$ . In Sec IV, we saw the effect of this showing up in the construction of the work probability distribution $\rho(W)$ from the moment generating function $K(t)$. If we use only the $n=1$ and $n=2$ terms of the infinite series in Eq. ( 4.1 ), we get the Gaussian probability distribution shown in Eq. (4. 16a). If we want to calculate the tail of $\rho(W)$, then we need a closed form for $K(t)$ which can be obtained from the large-$n$ terms of Eq. (4.1 ). The average value of $W$ that one infers from this exercise is necessarily of $O(V / L)$ . This Appendix describes how one interpolates the probability distribution from the very small values of $W$ to values of the $O(V / L)$ and to even larger ones.

In the case of short- ranged correlations, as shown in Eq. (1.3), the average $W$ is of $O(N)$ which is the same as $O(V)$ and the distribution is centred round this average. The distribution is also very sharp and a departure of $O(1)$ from the mean results in a suppression of the probability by the factor $O(\exp(-N))$ . This is what causes the large value of $\varepsilon_\Omega$, the fluctuation in $\exp(-W)$ which is almost negligible at the peak of the distribution .

For the long-ranged correlations we see that the distribution for $W \to 0$ is given by the structure (see Eq. (4.18a)), $\rho(W) \propto \exp\left(-\frac{L_\perp^2}{L^2}\left[\ln\frac{\bar{W}}{W}\right]^2\right)$, where the scale $\bar{W}$ is found to be proportional to $L_\perp^2$ ($O(V/L)$) and the overall scale of the distribution is $L_\perp^2/L^2 \propto V/L^3$. This form holds for $W \ll \bar{W}$ and produces a long-lived tail compared to the short-ranged case where the prefactor of the logarithm is proportional to $V$. We note that if $\bar{W}$ and $W$ are of the same order, then the magnitude $\rho(W)$ is $O(\exp-(\frac{L_\perp^2}{L^2}))$. This is the same kind of probability distribution that one obtains from the $\rho(W)$ of Eq. (4.18b). Thus the low-$W$ form of Eq. (4.18a) merges into the Gaussian of Eq. (4.18b). However, using the Gaussian of Eq. (4.16a) which gives the average of $W$ correctly, we get a similar order of $\rho(W)$ only when $W - \langle W \rangle \approx 1/L$.

Thus, the picture that emerges is that one has a Gaussian distribution which actually takes note of the fact that $\langle W \rangle$ is $O(V)$ and we can write this distribution as $\rho(W) \propto \exp(-L_\perp^2(\frac{W}{<W>}-1)^2)$ in the vicinity of $\langle W \rangle$. Clearly, if we consider values of $\frac{W}{\langle W \rangle}-1$ which are $O(1/L)$, then the probability coming from such a distribution merges with the values obtained from the distribution found in Eq. (4.16) which in turn merges with the small $W$ tail (as also the large $W$ tail) in the appropriate limit. Given this, if we write the work probability distribution as

$$\rho(W) \propto e^{-L_\perp^2\left(\frac{W}{\langle W \rangle}-1\right)^2} \qquad (A1)$$

then

$$\langle W_{fl}^2 \rangle = \langle (W - \langle W \rangle)^2 \rangle = \langle W \rangle^2 / 2L_\perp^2 \qquad (A2)$$

Setting $\langle W \rangle = \sigma L_\perp^2 L$, where $\sigma$ is a number of O(1), $\langle W_{fl}^2 \rangle = \sigma^2 L_\perp^2 L^2/2$ and comparing with the exact answer of Eq. (3.14) we find that

$$\sigma^2 = \left(\frac{h}{MJ}\right)^2 \frac{1}{3\pi} \qquad (A3)$$

We can now evaluate the moments $\langle e^{-nW} \rangle$ and find

$$\varepsilon_\Omega = e^{\frac{1}{6\pi} \frac{L_\perp^2}{L^2}\left(\frac{hL^2}{MJ}\right)^2} = e^{\frac{1}{6\pi} AL_h^4} \qquad (A4)$$

in exact accordance with what was reported in Ref [23] and obtained in Eq. (4.21). This is also of the same form as in Eq. (4.24), the only difference being in the prefactor of $AL_h^4$.

As for $W \gg \langle W \rangle$, we note that the tail expressed by Eq. (4.18c) gives a $\rho(W)$ of $O(\exp-\left(\frac{L_\perp^2}{L}\right))$. The distribution $\rho(W)$ of Eq. (A1) gives a result of same order for $\frac{W}{\langle W \rangle} - 1 \approx O\left(\frac{1}{\sqrt{L}}\right)$. The above discussion supports the following picture for the work distribution function:

i) It is centred at $W$ of $O(\langle W \rangle)$ with $\langle W \rangle$ proportional to the volume of the system as shown in Eq. (A1) and its width is proportional to $L_\perp^2 L^2$.

ii) The tail of the distribution ($W \to 0$ i.e. $W \ll L_\perp^2$) is proportional to $\exp\left(-A\left(\ln\frac{1}{W}\right)^2\right)$ and the distribution of Eq. (A1) merges into this tail for $\langle W \rangle - W > \frac{1}{L}$

iii) For $W - \langle W \rangle > O(\frac{1}{\sqrt{L}})$, the distribution of Eq. (A1) merges with the large W tail of Eq. (4.18c)

The weakening of the mean square fluctuation is due to the presence of the prolonged $W \approx 0$ tail in the distribution function.